\documentclass[twoside,twocolumn,9pt]{article}
\usepackage{extsizes}
\usepackage[super,sort&compress,comma]{natbib} 
\usepackage[version=3]{mhchem}
\usepackage[left=1.5cm, right=1.5cm, top=1.785cm, bottom=2.0cm]{geometry}
\usepackage{balance}
\usepackage{times,mathptmx}
\usepackage{sectsty}
\usepackage{graphicx} 
\usepackage{lastpage}
\usepackage[format=plain,justification=justified,singlelinecheck=false,font={stretch=1.125,small,sf},labelfont=bf,labelsep=space]{caption}
\usepackage{float}
\usepackage{fancyhdr}
\usepackage{fnpos}
\usepackage[english]{babel}
\usepackage{array}
\usepackage{droidsans}
\usepackage{charter}
\usepackage[T1]{fontenc}
\usepackage[usenames,dvipsnames]{xcolor}
\usepackage{setspace}
\usepackage[compact]{titlesec}
\usepackage{gensymb}

\newcommand{\onlinecite}[1]{\hspace{-1 ex} \nocite{#1}\citenum{#1}}

\definecolor{cream}{RGB}{222,217,201}

\begin{document}

\pagestyle{fancy}
\thispagestyle{plain}
\fancypagestyle{plain}{

}

\makeFNbottom
\makeatletter
\renewcommand\LARGE{\@setfontsize\LARGE{15pt}{17}}
\renewcommand\Large{\@setfontsize\Large{12pt}{14}}
\renewcommand\large{\@setfontsize\large{10pt}{12}}
\renewcommand\footnotesize{\@setfontsize\footnotesize{7pt}{10}}
\makeatother

\renewcommand{\thefootnote}{\fnsymbol{footnote}}
\renewcommand\footnoterule{\vspace*{1pt}%
\color{cream}\hrule width 3.5in height 0.4pt \color{black}\vspace*{5pt}} 
\setcounter{secnumdepth}{5}

\makeatletter 
\renewcommand\@biblabel[1]{#1}            
\renewcommand\@makefntext[1]%
{\noindent\makebox[0pt][r]{\@thefnmark\,}#1}
\makeatother 
\renewcommand{\figurename}{\small{Fig.}~}
\sectionfont{\sffamily\Large}
\subsectionfont{\normalsize}
\subsubsectionfont{\bf}
\setstretch{1.125} 
\setlength{\skip\footins}{0.8cm}
\setlength{\footnotesep}{0.25cm}
\setlength{\jot}{10pt}
\titlespacing*{\section}{0pt}{4pt}{4pt}
\titlespacing*{\subsection}{0pt}{15pt}{1pt}

\fancyfoot{}
\fancyfoot[RO]{\footnotesize{\sffamily{1--\pageref{LastPage} ~\textbar  \hspace{2pt}\thepage}}}
\fancyfoot[LE]{\footnotesize{\sffamily{\thepage~\textbar\hspace{3.45cm} 1--\pageref{LastPage}}}}
\fancyhead{}
\renewcommand{\headrulewidth}{0pt} 
\renewcommand{\footrulewidth}{0pt}
\setlength{\arrayrulewidth}{1pt}
\setlength{\columnsep}{6.5mm}
\setlength\bibsep{1pt}

\makeatletter 
\newlength{\figrulesep} 
\setlength{\figrulesep}{0.5\textfloatsep} 

\newcommand{\topfigrule}{\vspace*{-1pt}%
\noindent{\color{cream}\rule[-\figrulesep]{\columnwidth}{1.5pt}} }

\newcommand{\botfigrule}{\vspace*{-2pt}%
\noindent{\color{cream}\rule[\figrulesep]{\columnwidth}{1.5pt}} }

\newcommand{\dblfigrule}{\vspace*{-1pt}%
\noindent{\color{cream}\rule[-\figrulesep]{\textwidth}{1.5pt}} }

\makeatother

\twocolumn[
  \begin{@twocolumnfalse}
\vspace{3cm}
\sffamily
\begin{tabular}{m{4.5cm} p{13.5cm} }

& \noindent\LARGE{\textbf{Nonspherical armoured bubble vibration$^\dag$}} \\
\vspace{0.3cm} & \vspace{0.3cm} \\

 & \noindent\large{\textbf{G. Prabhudesai, I. Bihi, F. Zoueshtiagh, J. Jose,  and M. Baudoin \textit{$^{\ast}$}}} \\

 & \noindent\normalsize{In this paper, we study the dynamics of cylindrical armoured bubbles excited by mechanical vibrations. A step by step transition from cylindrical to spherical shape is reported as the intensity of the vibration is increased, leading to a reduction of the bubble surface and a  dissemination of the excess particles. We demonstrate through energy balance that nonspherical armoured bubbles constitute a metastable state. The vibration instills the activation energy necessary for the bubble to return to its least energetic stable state: a spherical armoured bubble. At this point, particle desorption can only be achieved through higher amplitude of excitation required to overcome capillary retention forces. Nonspherical armoured bubbles open perspectives for tailored localized particle dissemination with limited excitation power.
} \\

\end{tabular}

 \end{@twocolumnfalse} \vspace{0.6cm}

  ]

\renewcommand*\rmdefault{bch}\normalfont\upshape
\rmfamily
\section*{}
\vspace{-1cm}


\footnotetext{\textit{Univ. Lille, CNRS, Centrale Lille, ISEN, Univ. Valenciennes, UMR 8520, International Laboratory LEMAC/LICS - IEMN, F-59000 Lille, France}}
\footnotetext{$\ast$ Corresponding author. E-mail: michael.baudoin@univ-lille1.fr}

\footnotetext{\dag~Electronic Supplementary Information (ESI) available: [details of any supplementary information available should be included here]. See DOI: 10.1039/b000000x/}



\section{Introduction}

Armoured bubbles, i.e. bubbles coated with a dense layer of partially wetting particles, exhibit fascinating properties such as increased stability toward dissolution \cite{l_du_2003,prl_abkarian_2007} or the ability to sustain non-spherical shapes \cite{nature_subramaniam_2005}. These bubbles can be massively produced by bulk emulsification techniques \cite{nm_binks_2006,cup_binks_2006,jacs_fuji_2006} or by bubble injection in a suspension of particles \cite{jmc_studart_2007}, but with a limited control over the bubbles properties. Recently microfluidic techniques have been considered to produce some calibrated armoured bubbles of tailored shapes, sizes and composition \cite{sm_brugarolas_2013}. Some of these techniques simply rely on the natural inclination of particles to be adsorbed at air/liquid interfaces \cite{naturec_subramanian_2005,sm_kotula_2012}, while others exploit fancy chemical- or temperature-mediated processes \cite{a_park_2009,c_tumarkin_2011} or the deep modification of liquid/air interface dynamics in particle covered capillary tubes \cite{sm_zoueshtiagh_2014,prl_bihi_2016}. In parallel, much effort has been devoted to the characterization of armoured bubbles properties, and in particular their stability toward dissolution \cite{l_du_2003,prl_abkarian_2007}, their lifetime when exposed to surfactant \cite{l_subramaniam_2006} or their mechanical strength when submitted to ambient overpressure \cite{prx_taccoen_2016}.  Surprisingly, less effort has been devoted to the study of the response of armoured bubbles to mechanical vibrations \cite{pnas_poulichet_2015,sm_poulichet_2016}, while bubbles are known to be outstanding resonators \cite{pm_strutt_1917,pm_minnaert_1933,asme_plesset_1949,jap_keller_1956,oup_brennen_1995}, with high quality factors and remarkably low resonance frequencies \cite{pm_minnaert_1933}. Recently, Poulichet et al. \cite{pnas_poulichet_2015,sm_poulichet_2016} investigated the acoustically driven oscillation of particle-coated bubbles. They showed that these oscillations can lead to the tailored expulsion of the particles from the surface at sufficient acoustic power. To the best our knowledge, the response of their nonspherical counterpart to mechanical vibrations has not been reported so far in the literature.

In this paper, we characterize  the dynamics of cylindrical armoured bubbles excited with vertical vibrations produced by a vibration exciter. We show that the vibration leads to a step by step transition of the bubble shape from cylindrical to spherical as the amplitude of the vibration is increased. This shape evolution leads to a decrease of the surface of the armoured bubble and thus a release of excess particles in the liquid.  This evolution is shown to be highly dependent on the excitation frequency: lower excitation amplitude is required near the Minnaert bubble resonance to trigger this transition. We demonstrate through energy balance that nonspherical armoured bubbles constitute a metastable state. The bubble vibration enables the release of the mechanical constraint between the particles which maintains the bubble shape and thus lets it evolve toward the least energetic stable state: a spherical armoured bubble. This transition does not require to overcome capillary retention forces and thus enables particle dissemination with reduced excitation power. Once the armoured bubble is spherical, massive particle desorption is only observed when the kinetic energy instilled from the surface vibration overcomes the capillary retention forces. The required power in this case is thus larger than for nonspherical bubble and should increase as the particle dimensions are decreased as demonstrated through dimensional analysis. 

\section{Materials and methods}

Cylindrical armoured bubbles of tailored radius $R_c = 0.49 \pm 0.01$ mm and length $L_c = 8.5 \pm 0.5$ mm are produced with the method introduced in ref.\onlinecite{sm_zoueshtiagh_2014}:  First, glass capillary tubes of radius $R_t = 501$ $\mu$m are cleaned by successive sonication in acetone and alcohol. Then the tubes are treated with piranha solution (sulfuric acid + hydrogen peroxide) to clean organic residues off the glass. They are then kept in DI Water in a sealed recipient before use. Prior to the particle dispersion, they are dried for $1$ hour in an oven at $120 \degree C$. Second, Rilsan (Polyamide 11) particles of mean radius $R_p = 15 \pm 1$ $\mu$m are scattered into the tube by gently blowing them with an air jet. Third, a DI water liquid finger is pushed at constant flow rate ($Q = 0.2$ ml/h) with a syringe pump inside the tube leading to the formation of a long cylindrical armoured bubble through the process described in ref.\onlinecite{sm_zoueshtiagh_2014}. 

Finally the bubble is extracted from the tube  and  injected inside a rectangular plexiglas closed chamber of height $20$ mm, length $75$ mm and width $55$ mm filled with DI Water, by applying a larger flow rate of $Q = 5$ ml/h. The bubbles are cut at the appropriate length  by twisting the capillary sideways while keeping it horizontal. This cuts the bubble on the open end of the capillary.

\begin{figure}[htbp]
\centering
  \includegraphics[width=0.5 \textwidth]{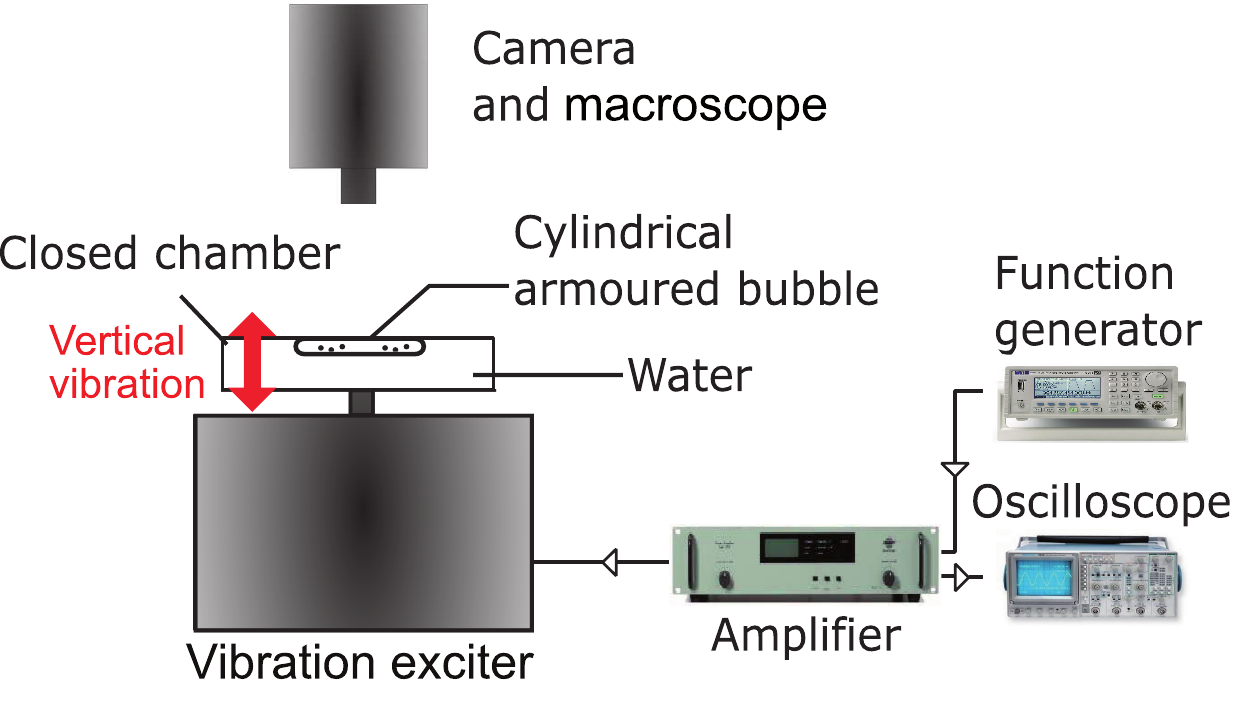}
  \caption{Scheme of the experimental setup}
  \label{fig:sketch}
\end{figure}

Then the chamber is vibrated vertically with a Bruel \& Kjaer 4809 vibration exciter driven by a sinusoidal Voltcraft FG 250D function generator whose signal is amplified with a 2718 Bruel \& Kjaer amplifier. The amplitude of the excitation $A$ is measured with an IEPE Deltatron type 4519-002 Bruel \& Kjaer accelerometer. The evolution of the bubble shape is recorded with a Hamamatsu C9300 high resolution camera mounted on an Olympus SZX7 macroscope, while the fast bubble oscillations are captured with a Photron SA3 high speed camera. Finally images are treated with ImageJ software and the evolution of the bubble shape is quantified by using the roundness shape descriptor $R_o = 4  S /  \pi L_m^2$, with S the apparent surface of the bubble (in the pictures), and $L_m$ the length of the bubble major axis. This shape descriptor gives a measure of the deviation of the bubble from a circular shape, with $R_o = 1$ as the bubble becomes perfectly spherical.

\section{Experimental results}

\subsection{Shape evolution}

\begin{figure}[htbp]
\centering
  \includegraphics[width=0.5 \textwidth]{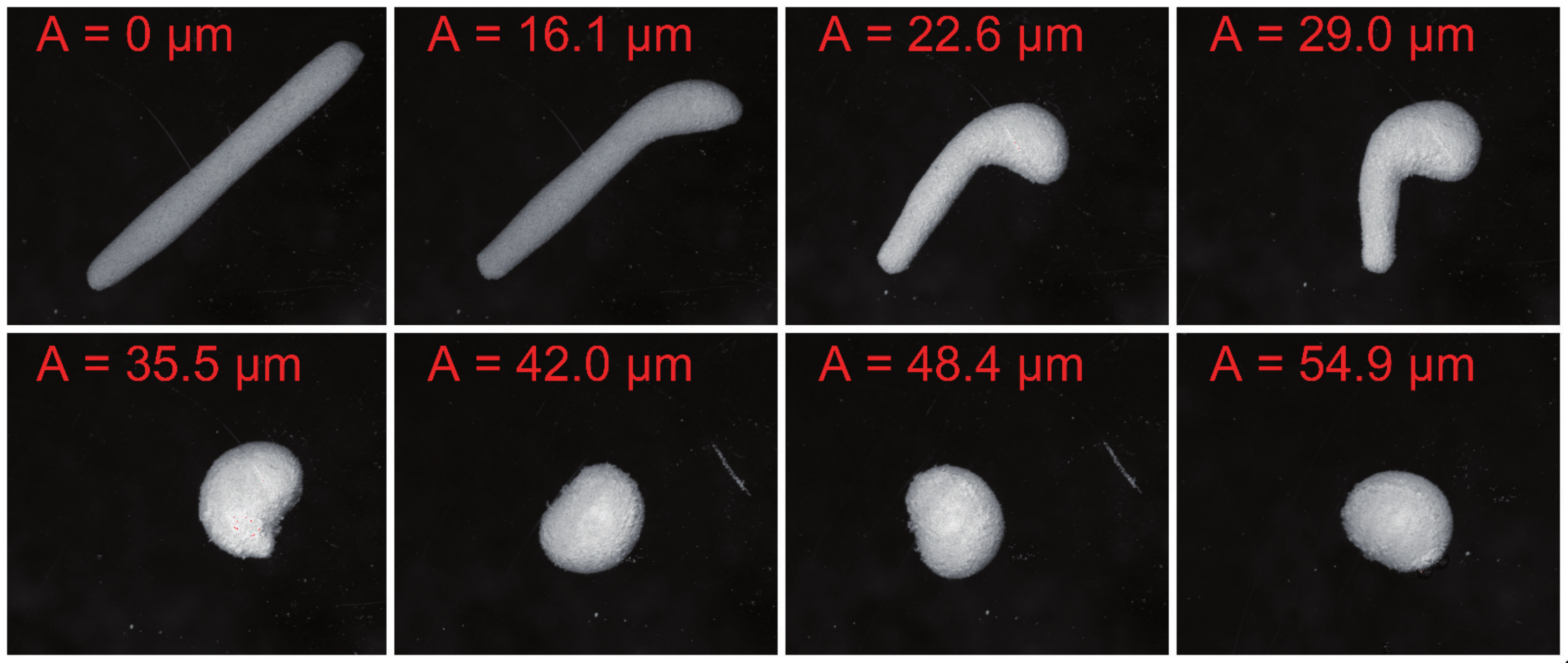}
  \caption{Evolution of a cylindrical armoured bubble of length $L_c = 8.5$ mm and radius $R_c = 0.5$ mm vibrated vertically with a vibration exciter at frequency $f_e = 1$ kHz and at an amplitude $A$ indicated at the top of each picture. Each picture is taken after two minutes of excitation.}
  \label{fig:pict1}
\end{figure}

\begin{figure}[htbp]
\centering
  \includegraphics[width=0.5 \textwidth]{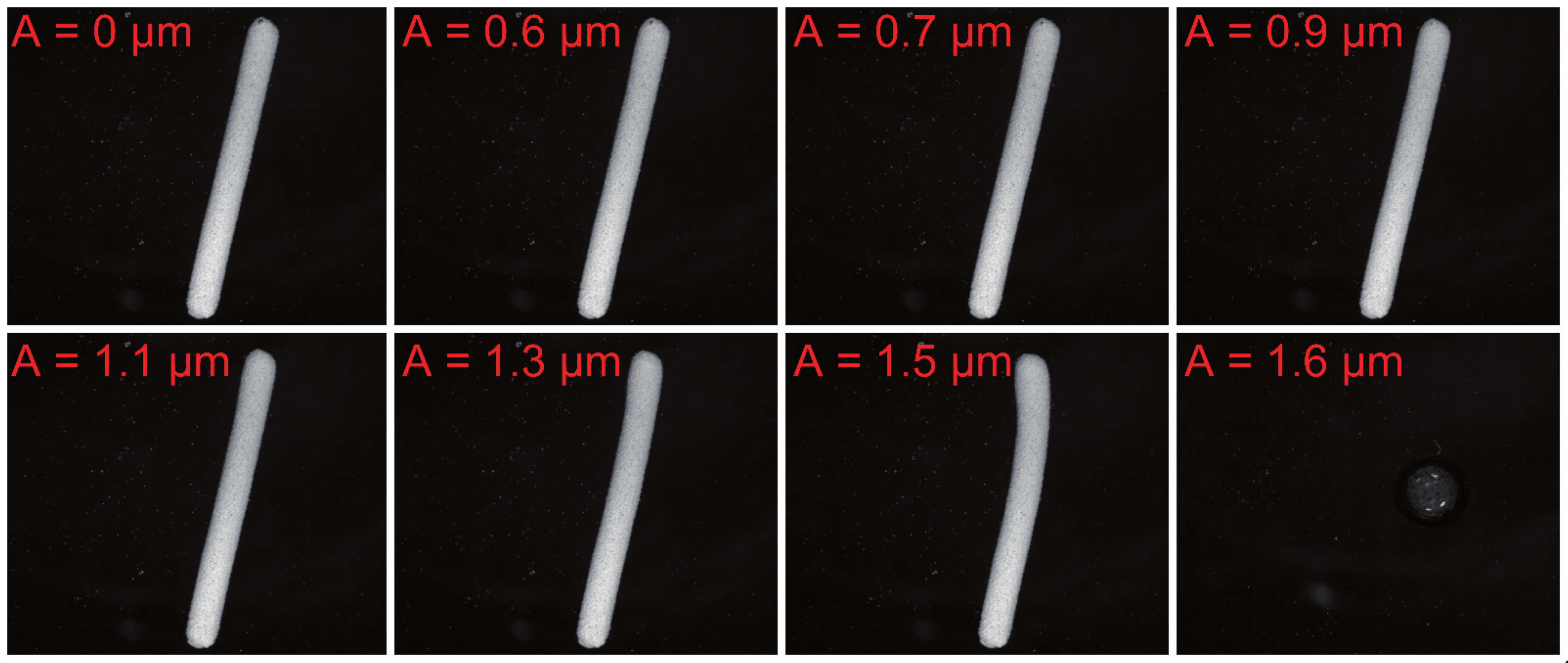}
  \caption{Evolution of a cylindrical armoured bubble of length $L_c = 8.5$ mm and radius $R_c = 0.5$ mm vibrated vertically  with a vibration exciter at frequency $f_e = 2.5$ kHz and at an amplitude $A$ indicated at the top of each picture. Each picture is taken after two minutes of excitation. In the final picture, all the particles initially at the surface of the bubble have been disseminated in the liquid.}
  \label{fig:pict2p5}
\end{figure}

\begin{figure}[htbp]
\centering
  \includegraphics[width=0.5 \textwidth]{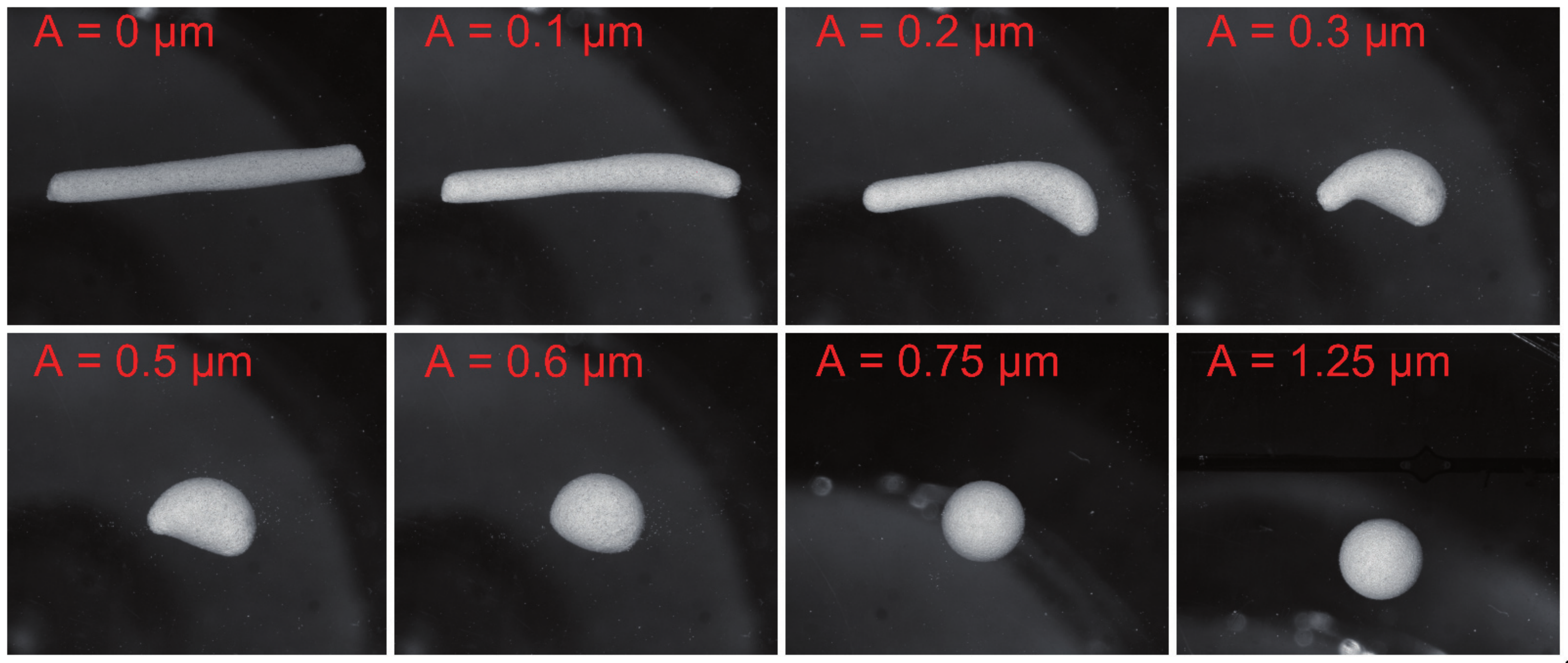}
  \caption{Evolution of a cylindrical armoured bubble  of length $L_c = 8.5$ mm and radius $R_c = 0.5$ mm vibrated vertically with a vibration exciter at frequency $f_e = 3$ kHz and at an amplitude $A$ indicated at the top of each picture. Each picture is taken after two minutes of excitation.}
  \label{fig:pict3}
\end{figure}

\begin{figure}[htbp]
\centering
  \includegraphics[width=0.5 \textwidth]{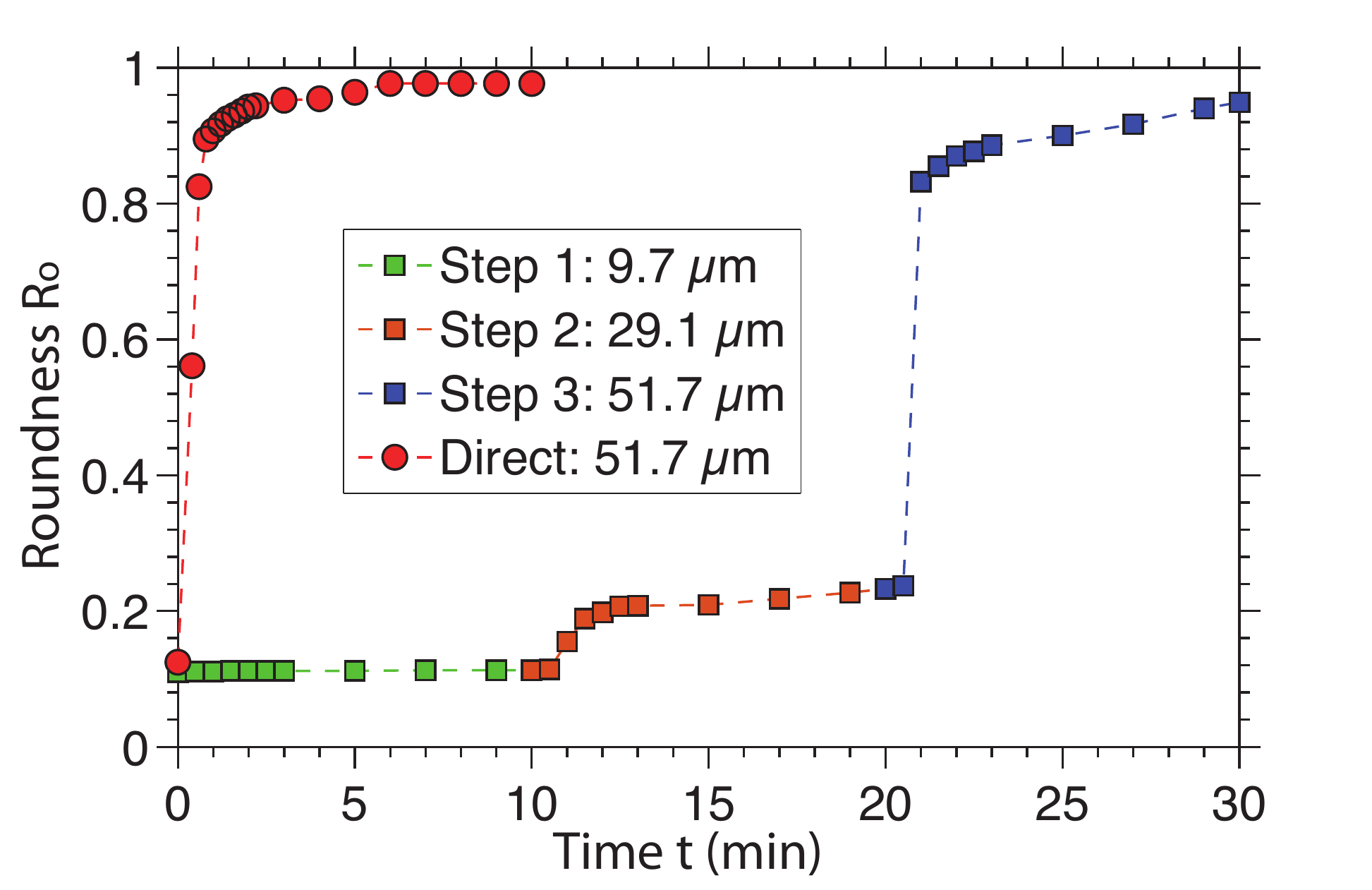}
  \caption{Transient evolution of the roundness $Ro$ of a cylindrical armoured bubble excited at $1$ kHz as  function of time (in min). The squares show the evolution of the bubble roundness when the amplitude $A$ is increased step by step every $10$ min from $A = 9.7 \; \mu$m to $A=51.7 \; \mu$m. The circles show the evolution when the amplitude is directly set up to $51.7 \; \mu$m.}
  \label{fig:transient}
\end{figure}

\begin{figure}[htbp]
\centering
  \includegraphics[width=0.5 \textwidth]{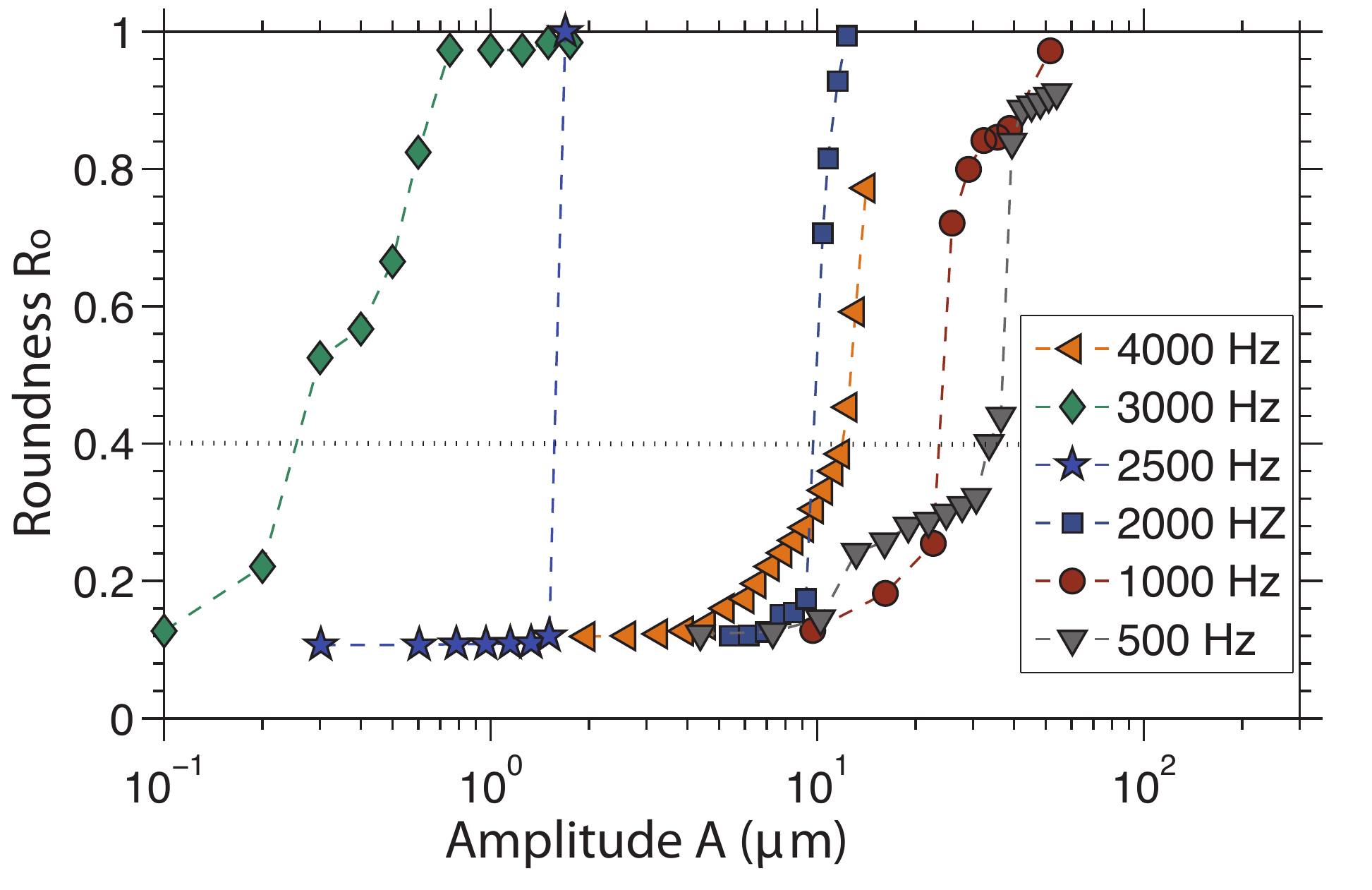}
  \caption{Evolution of the Roundness  as a function of the excitation amplitude for different excitation frequencies $f_e$. For each frequency, the roundness is measured sequentially after 2 minutes of excitation at each amplitude.}
  \label{fig:amplitude}
\end{figure}

The vertical vibration of the cylindrical armoured bubbles at a given amplitude A ($0.1 \; \mu$m$<A<50 \; \mu$m) and frequency $f_e$ ($0.5$kHz$<f_e<4$kHz) leads to an isovolume evolution of its shape through a reduction of its surface  and a release of excess particles in the liquid (see Fig. \ref{fig:pict1} to \ref{fig:pict3} and Movie S1). When the power is turned on, the bubble shape evolves at first over characteristic times $<1min$ and then evolves slowly toward an asymptotic shape, which depends on the excitation amplitude $A$ and the excitation frequency $f_e$ (see Fig. \ref{fig:transient}). As the amplitude of excitation is increased gradually, this asymptotic shape is more and more spherical, eventually leading to a spherical armoured bubble ($Ro \rightarrow 1$) (see Fig. \ref{fig:pict1} and \ref{fig:pict3}).
This transition is nevertheless not regular as a function of the excitation amplitude $A$: Under a frequency-dependent threshold, the bubble keeps its cylindrical shape (see Fig. \ref{fig:transient} and \ref{fig:amplitude}). Above this threshold, the bubble shape evolves at first weakly as the amplitude of the excitation is increased and then undergoes a sharp transition toward a spherical shape for a critical amplitude $\approx A_c$, defined here as the amplitude necessary for the bubble to achieve a roundness of $0.4$ (see Fig. \ref{fig:amplitude}).

\subsection{Minnaert resonance frequency}

\begin{figure}[htbp]
\centering
  \includegraphics[width=0.5 \textwidth]{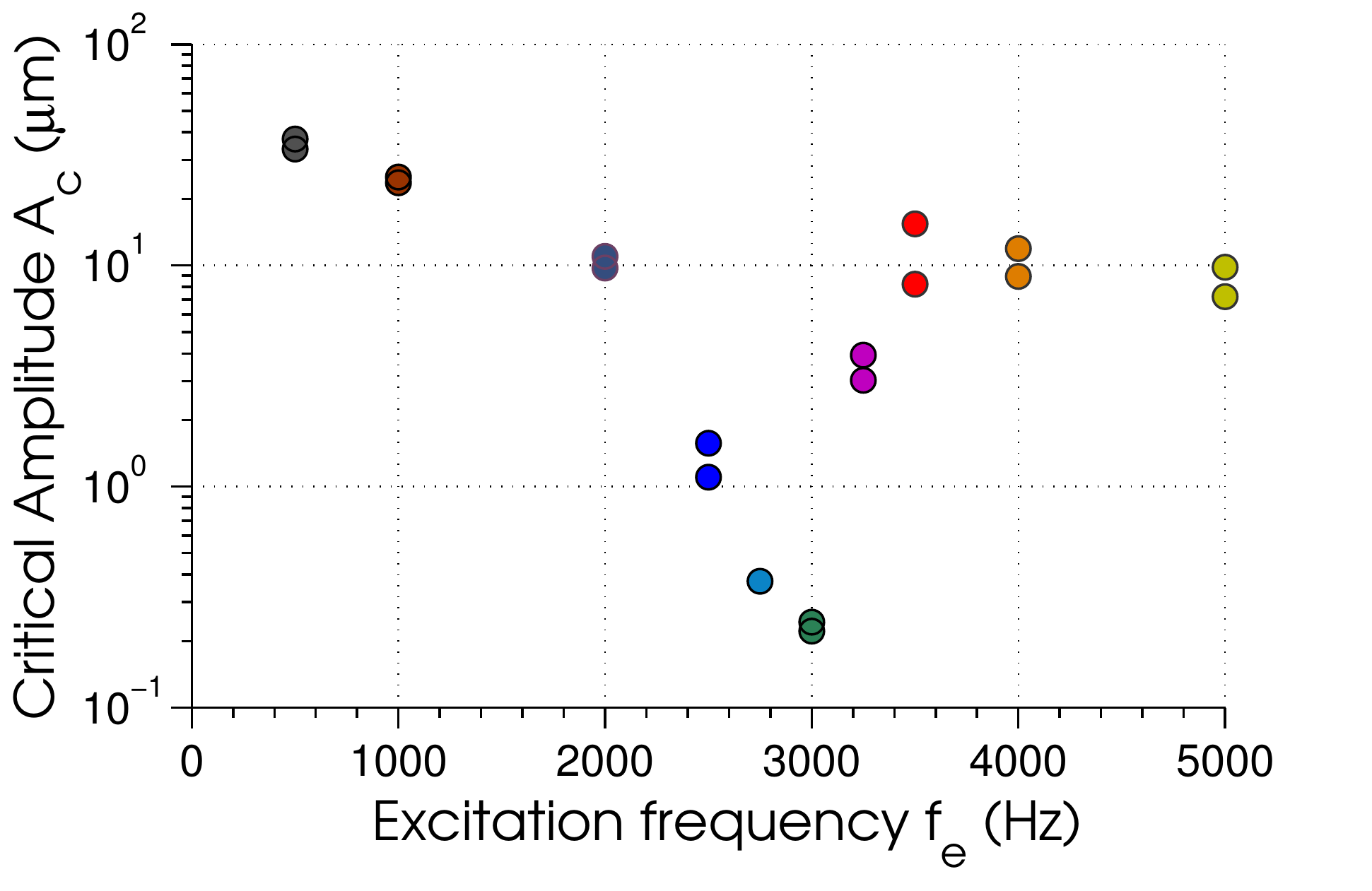}
  \caption{Critical excitation amplitude $A_c$ required for the bubble to reach a roundness $R_o = 0.4$ as a function of the excitation frequency $f_e$. Two experiments have been carried out for each frequency.}
  \label{fig:fe}
\end{figure}

This critical amplitude  $A_c$ highly depends on the excitation frequency $f_e$ (see Fig. \ref{fig:fe}) and decreases by several orders of magnitude close to the Minnaert resonance frequency of the bubble $f_M  = 3.26 \times R_s^{-1}$, with $R_s = \left( 3/4 \pi \; V \right)^{-1/3}$ the radius of a spherical bubble of same volume $V = \pi R_c^2 L_c$ as the initial cylindrical armoured bubble. Indeed we verified that the bubble undergoes a isovolume shape transformation (less than $4\%$ of volume change have been measured between the initial and final state, including errors in the estimation of the volume from the 2D pictures of the bubble). The precedent formula gives a value of the Minnaert resonance frequency of $f_M = 2.8\pm0.1$ kHz. 

It is interesting to note that two extremely different behaviors are observed around the Minnaert frequency: At $3$ kHz the transition from cylindrical to spherical shape is achieved with the weakest excitation amplitude. The evolution of the bubble shape starts at very low excitation amplitude ($A = 0.1 \; \mu$m $\ll A_c$) and then goes on progressively when the amplitude is increased, until a spherical armoured bubble is obtained (see Fig. \ref{fig:pict3} and Fig. \ref{fig:amplitude}). At $2.5$ kHz, no shape evolution is observed until the critical amplitude $A_c$ is reached. Then, when $A=A_c$ the bubble directly evolves into a particle-free spherical bubble with a release of all surrounding particles in the liquid (see Fig. \ref{fig:pict2p5}, Fig. \ref{fig:amplitude} and movie S2). 

\section{Discussion}


\subsection{Cylindrical bubbles: a metastable state}

\begin{figure}[htbp]
\centering
  \includegraphics[width=0.5 \textwidth]{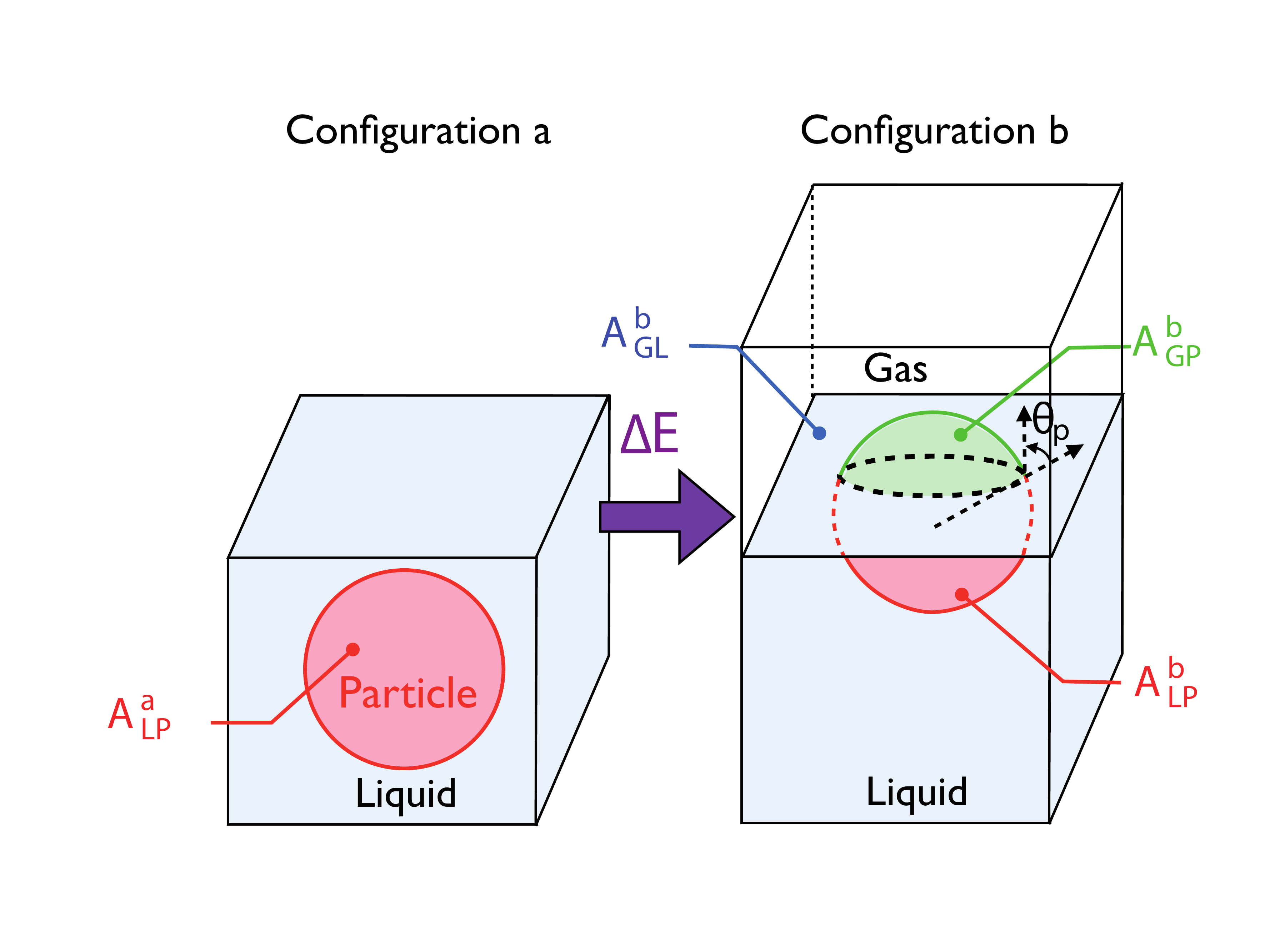}
  \caption{Sketch illustrating the migration of a particle lying in the liquid to an extended meniscus.}
  \label{fig:energy}
\end{figure}

Nonspherical armoured bubbles constitute a metastable state. This can be shown by computing the interfacial energy required for a quasi-static and isovolume transformation of a spherical armoured bubble into a nonspherical one. Such a transformation necessarily requires an increase in the particles-covered interface of the bubble since the spherical shape minimizes the bubble surface. We can compute the energy $\Delta E$, which is necessary for the migration of a particle from the liquid to an extended air-liquid interface (see Fig. \ref{fig:energy}). In the calculation below we assume that the particles are hydrophilic (with a liquid-air-particle contact angle $0\leq\theta_p\leq\pi/2$), spherical and that their radius $R_p$ is much smaller than the radius of curvature of the bubble.

The energy $\Delta E$ is simply the difference between the interfacial energy $E_b$ of the particle integrated at the air-liquid interface and the interfacial energy $E_a$ of the particle lying in the liquid:
\begin{equation}
\Delta E = E_b - E_a
\label{eq:delta}
\end{equation}
The interfacial energy $E_a$ in configuration (a) is:
\begin{equation}
E_a = \gamma_{LP} A_{LP}^a
\end{equation}
with $\gamma_{LP}$ the interfacial energy per unit area of the liquid-particle interface and $A_{LP} = 4 \, \pi \, R_p^2$ the liquid-particle interface in the configuration (a). The interfacial energy $E_b$ in the configuration (b) is the sum of the  gas-particle (GP), liquid-particle (LP) and gas-liquid (GL) interfacial energies:
\begin{equation}
E_b = \gamma_{GP} A_{GP}^b + \gamma_{LP} A_{LP}^b + \gamma_{GL} A_{GL}^b 
\end{equation}
with $\gamma$ and $A$ the interfacial energies per unit area and the areas of the corresponding interfaces. Since the particles intersect the liquid-air interface with a contact angle $\theta_P$ (to minimize interfacial energy), we deduce from simple geometric considerations that:
\begin{eqnarray}
& & A_{GP}^b = 2 \pi R_p^2 (1 - \cos \theta_P) \\
& & A_{LP}^b = 2 \pi R_p^2 (1 + \cos \theta_P) 
\end{eqnarray}
Then, due to their shape and arrangement, the particles cover only a fraction $\phi$ of the gas-liquid interface, called the specific surface area. Thus, the gas-liquid interface that is necessary to incorporate a new particle is $\pi R_P^2 / \phi$, with $\phi < 1$. If we subtract the surface occupied by the particle itself, we obtain:
\begin{equation}
A_{GL}^b = \pi R_P^2 / \phi - \pi R_P^2 \sin^2 \theta_P.
\label{eq:agl}
\end{equation}
If we now combine equations \ref{eq:delta} to \ref{eq:agl}, we obtain:
\begin{eqnarray}
\Delta E & = &  \pi R_p^2 \left[ 2 \gamma_{GP}  (1 - \cos \theta_P) + 2 \gamma_{LP}  (1 + \cos \theta_P) \right. \nonumber \\
& & \left. + \gamma_{GL} \left(1 / \phi -  \sin^2(\theta_P) \right)  \right]   - \left[ \gamma_{LP} 4 \, \pi \, R_p^2 \right] \nonumber
\end{eqnarray}
From the definition of the contact angle $\cos \theta_P = (\gamma_{GP} - \gamma_{LP}) / \gamma_{GL}$, we finally obtain:
\begin{equation}
\Delta E =  \gamma_{GL} \pi R_P^2 \left[ \frac{1}{\phi} - (1 - \cos \theta_P)^2   \right]
\end{equation}
Since $0 \leq (1-\cos \theta_P)^2 \leq 1$ (for hydrophilic particles) and $1 / \phi > 1$, we have\footnote{NB: The calculation of the energy required for the migration of a hydrophobic particle from the gas to the air-liquid interface, also gives $\Delta E > 0$}:
$$
\Delta E > 0
$$
This energy is positive indicating that the quasi-static transformation of a spherical bubble into a nonspherical one costs some energy. In other words a nonspherical bubble is not a stable state and would return to its least energetic state, i.e. a spherical armoured bubble in the absence of other constraint. Nevertheless, as discussed initially in ref.\onlinecite{nature_subramaniam_2005}, the jamming between the particles prevents this evolution and maintain the bubble in a metastable state. 

In the present experiments, the bubble vibration instills the activation energy required to induce this transition. Indeed, during the expansion phase of the bubble oscillation, the particles are no more in contact, and are thus freed from their interparticle mechanical constraint. In addition, the kinetic energy instilled to the particles by the vibration of the surface contribute to the release of the particles. This enables the bubble shape reconfiguration into a spherical armoured bubble.

\subsection{Particle expulsion from spherical armoured bubbles}

Spherical armoured bubbles on the other hand constitute a stable state. Indeed, a bubble without particles is a minimum surface energy configuration. Then, if a particle lying in the liquid migrates on the already existing liquid-air interface, the energy necessary for this migration $\Delta E^*$ is simply \cite{cocis_binks_2002}:
$$
\Delta E^* = \Delta E - \frac{\pi R_P^2}{\phi} \gamma_{GL} = - \gamma_{GL} \pi R_P^2 \ (1 - \cos \theta_P)^2 < 0
$$
which is negative, indicating that the minimization of surface energy leads to an adsorption of the particles which come into contact with the air-liquid interface. This principle was used in ref.\onlinecite{naturec_subramanian_2005,sm_kotula_2012,prx_taccoen_2016} to create armoured bubbles. 

Thus the desorption of particles lying at the surface of a spherical armoured bubble requires to overcome capillary retention forces as noticed by Poulichet and Garbin \cite{pnas_poulichet_2015,sm_poulichet_2016}. These authors showed experimentally that localized particle desorption occurs for weak oscillations of spherical armoured bubbles but that the massive release of particles in the liquid requires higher amplitudes of vibrations. 

We can estimate the kinetic energy that is necessary to overcome the capillary barrier, by simply equating $E_c = \frac{1}{2} m_P V_p^2 $ and $|\Delta E^*|$, with $m_p = 4/3 \pi \rho_p R_p^3$ the mass of the particles, $\rho_p$ their density, $V_p = 2 \pi f_e \Delta R_s$ their velocity at the oscillating interface, and $\Delta R_s$ the radial variation of the bubble.

We performed experiments on individual spherical armoured bubbles of radius $R_s = 1.25$ mm covered with Rilsan particles of  radius $R_p \approx 15 \mu$m, density $\rho_p \approx 1050$ kg m$^{-3}$, contact angle $\theta_P \approx 70 \degree $ \cite{sm_zoueshtiagh_2014}, lying in water with surface tension $\gamma_{GL} \approx 70$ mN m$^{-1}$ (see e.g. movie S3). Massive particle detachment was observed when the bubble oscillation magnitude reaches $\Delta R_s / R_s \approx 10 \%$ of the initial radius of the bubble. This leads to $E_c / |\Delta E^*| \approx 0.4$. For the experiments performed at $2.5 kHz$ (see e.g. Movie S2), the ratio $E_c / |\Delta E^*|$ measured when massive desorption of the particles occurs is approximatively equal to $2 $. These experiments are thus consistant with this criterion.

It is also interesting to note that this simple criterion is in qualitative agreement with the data provided by Poulichet et al. \cite{sm_poulichet_2016}. Indeed, in their experiments, the latex particle size and density are respectively $R_p = 250$ nm and $\rho_p = 1040$ kg m$^{-3}$, the air-liquid-particle contact angle is $\theta_p \approx 45^o$ \cite{pnas_poulichet_2015}, the frequency of excitation is $f_e = 40$ kHz, the surface tension is $\gamma_{GL} \approx 70$ mN m$^{-1}$ and the amplitude of oscillation required for the particles dispersion is typically $50$ $\mu$m. This leads to a ratio $E_c / | \Delta E |$ required for particle dispersion of $\approx 3$.

\subsection{Analysis of the dynamics observed near the Minnaert resonance frequency}

The analysis detailed in the two previous subsection provides insightful elements, which help in understanding the two different behaviors observed at $2.5$ kHz and $3$ kHz, near the Minnaert resonance frequency of the spherical bubble. When the bubble is excited at $2.5$ kHz, no shape modification is observed when $A<A_c$ and then brutally, all particles are ejected from the interface when $A=A_c=1 \; \mu$m. While at $3$ kHz, the shape evolution starts at low amplitude $A=0.2 \; \mu$m$<<A_c$ and reaches a spherical armoured bubble shape at $A = 1 \; \mu$m. This suggests that $2.5$ kHz is closer to the resonance frequency of the spherical bubble and that $3$ kHz is closer to the resonance frequency of the cylindrical armoured bubble. Indeed, at $2.5$ kHz, the bubble response is weak (since the excitation differs from the cylindrical armoured bubble resonance frequency) until the vibration is sufficient to induce a shape transformation. As the shape becomes more and more spherical, the bubble amplitude increases up to an amplitude that is sufficient to overcome capillary retention forces, resulting in the release of all interfacial particles in the liquid. At $3$ kHz, the bubble starts responding at low amplitude of excitation (since the excitation frequency is close to the cylindrical armoured bubble resonance frequency), but as the bubble shape evolves, the excitation frequency differs from the resonance frequency, requiring more and more power to enable a shape evolution of the bubble. When the spherical shape is reached, the excitation amplitude of $A = 1 \; \mu$m is not sufficient to overcome the capillary retention forces and thus the bubble keeps its armour.

\subsection{Nonspherical bubbles: powerfull vectors for particles transport and dissemination with reduced activation power}

All the experimental and theoretical results provided in this paper indicate that particles dissemenation can be achieved with less power with nonspherical than with spherical armoured bubble. The dimensional analysis below suggests that this tendency should increase as the size of the bubble and particles are decreased. 

Let's consider a spherical bubble of radius $R_s$ covered with particles of radius $R_p$ excited at its resonance frequency $f_M$ (for maximum efficiency) oscillating at an amplitude of oscillation $\Delta \bar{R_s} = \Delta R_s / R_s$ sufficient to disseminate the particles. The equality $E_c = |\Delta E^*|$ leads to:
$$
1/2   \left[ \rho_p  \times (4/3 \pi R_p)^3 \right]  \times \left[ R_s \;  \Delta \bar{R_s} \times 3.26 \; R_s^{-1} \right]^2  =  \gamma_{GL} \pi R_P^2 \ (1 - \cos \theta_P)^2
$$
If the density and wetting properties of the particles are kept constant and only the size of the bubble and particles are changed, one obtain:
$$
 \Delta \bar{R_s} \propto R_p^{-1}
$$
The bubble amplitude of oscillation required for particles dissemination does not seem to depend on the bubble size but is nevertheless inversely proportional to the size of the particles. This means that for smaller particles, a larger relative amplitude of oscillation is required to untrap the particles located at the surface of a spherical armoured bubble. While for a cylindrical bubble, the same relative amplitude of oscillations would enable the release of particles since there is no need to overcome capillary retention forces, but only to suppress the contact between the particles. This might be essential for medical applications. Indeed, contrast agents currently used in the body have a typical size of a few microns and active material of typically a few nanometers are generally required for efficient dissemination in the body. It might thus become difficult to overcome capillary retention forces at this scale.

\section{Conclusion and perspectives}

In this paper, the response of nonspherical armoured bubbles to mechanical vibrations was studied experimentally and analyzed theoretically. In particular, it was shown that a nonspherical armoured bubble constitute a metastable state, which evolves toward its minimum energy state, i.e. a spherical armoured bubble, as its surface is vibrated. This shape evolution leads to a reduction of the bubble interface and consequently a release of the excess particles. Once the spherical shape is reached, the massive release of particles from the interface requires that the kinetic energy instilled to the particles by the vibration of the bubble surface exceeds the desorption energy. Scaling laws indicate that the release of particles from \textit{spherical} armoured bubbles becomes extremely challenging as the size of the bubbles is decreased. Thus nonspherical armoured bubbles appear as a tremendous alternative for controlled particle dissemination at small scales with limited input power. Applications can be envisioned in medicine for in-vivo active drug delivery through the use of robust acoustical methods precedently developed for ultrasound contrast agents.

\section*{Acknowledgements}
This work was supported by Agence Nationale de la Recherche (Grant No. ANR-12-BS09-0021- 01).


\begin{mcitethebibliography}{24}
\providecommand*{\natexlab}[1]{#1}
\providecommand*{\mciteSetBstSublistMode}[1]{}
\providecommand*{\mciteSetBstMaxWidthForm}[2]{}
\providecommand*{\mciteBstWouldAddEndPuncttrue}
  {\def\EndOfBibitem{\unskip.}}
\providecommand*{\mciteBstWouldAddEndPunctfalse}
  {\let\EndOfBibitem\relax}
\providecommand*{\mciteSetBstMidEndSepPunct}[3]{}
\providecommand*{\mciteSetBstSublistLabelBeginEnd}[3]{}
\providecommand*{\EndOfBibitem}{}
\mciteSetBstSublistMode{f}
\mciteSetBstMaxWidthForm{subitem}
{(\emph{\alph{mcitesubitemcount}})}
\mciteSetBstSublistLabelBeginEnd{\mcitemaxwidthsubitemform\space}
{\relax}{\relax}

\bibitem[Du \emph{et~al.}(2003)Du, Bilbao-Montoya, Binks, Dickinson, Ettelaie,
  and Murray]{l_du_2003}
Z.~Du, M.~Bilbao-Montoya, B.~Binks, E.~Dickinson, R.~Ettelaie and B.~Murray,
  \emph{Langmuir}, 2003,  3105--3108\relax
\mciteBstWouldAddEndPuncttrue
\mciteSetBstMidEndSepPunct{\mcitedefaultmidpunct}
{\mcitedefaultendpunct}{\mcitedefaultseppunct}\relax
\EndOfBibitem
\bibitem[Abkarian \emph{et~al.}(2007)Abkarian, Subramaniam, Kim, Larsen, Yang,
  and Stone]{prl_abkarian_2007}
M.~Abkarian, A.~Subramaniam, S.-H. Kim, R.~Larsen, S.-M. Yang and H.~Stone,
  \emph{Phys. Rev. Lett.}, 2007, \textbf{99}, 188301\relax
\mciteBstWouldAddEndPuncttrue
\mciteSetBstMidEndSepPunct{\mcitedefaultmidpunct}
{\mcitedefaultendpunct}{\mcitedefaultseppunct}\relax
\EndOfBibitem
\bibitem[Subramanian \emph{et~al.}(2005)Subramanian, Abkarian, Mahadevan, and
  Stone]{nature_subramaniam_2005}
A.~Subramanian, M.~Abkarian, L.~Mahadevan and H.~Stone, \emph{Nature}, 2005,
  \textbf{438}, 930\relax
\mciteBstWouldAddEndPuncttrue
\mciteSetBstMidEndSepPunct{\mcitedefaultmidpunct}
{\mcitedefaultendpunct}{\mcitedefaultseppunct}\relax
\EndOfBibitem
\bibitem[Binks and Murakami(2006)]{nm_binks_2006}
B.~Binks and R.~Murakami, \emph{Nat. Mater.}, 2006, \textbf{5}, 865--869\relax
\mciteBstWouldAddEndPuncttrue
\mciteSetBstMidEndSepPunct{\mcitedefaultmidpunct}
{\mcitedefaultendpunct}{\mcitedefaultseppunct}\relax
\EndOfBibitem
\bibitem[Binks and Horozov(2006)]{cup_binks_2006}
B.~Binks and T.~Horozov, \emph{Colloidal particles at liquid interfaces},
  Cambridge University Press, 2006\relax
\mciteBstWouldAddEndPuncttrue
\mciteSetBstMidEndSepPunct{\mcitedefaultmidpunct}
{\mcitedefaultendpunct}{\mcitedefaultseppunct}\relax
\EndOfBibitem
\bibitem[Fujii \emph{et~al.}(2006)Fujii, Ryan, and Armes]{jacs_fuji_2006}
S.~Fujii, A.~Ryan and S.~Armes, \emph{J. Am. Chem. Soc.}, 2006, \textbf{128},
  7882--7886\relax
\mciteBstWouldAddEndPuncttrue
\mciteSetBstMidEndSepPunct{\mcitedefaultmidpunct}
{\mcitedefaultendpunct}{\mcitedefaultseppunct}\relax
\EndOfBibitem
\bibitem[Studart \emph{et~al.}(2007)Studart, Gonzenbach, Akartuna, Tervoort,
  and Gauckler]{jmc_studart_2007}
A.~Studart, U.~Gonzenbach, I.~Akartuna, E.~Tervoort and L.~Gauckler, \emph{J.
  Mater. Chem.}, 2007, \textbf{17}, 3283--3289\relax
\mciteBstWouldAddEndPuncttrue
\mciteSetBstMidEndSepPunct{\mcitedefaultmidpunct}
{\mcitedefaultendpunct}{\mcitedefaultseppunct}\relax
\EndOfBibitem
\bibitem[Brugarolas \emph{et~al.}(2013)Brugarolas, Tu, and
  Daeyeon]{sm_brugarolas_2013}
T.~Brugarolas, F.~Tu and L.~Daeyeon, \emph{Soft Matter}, 2013, \textbf{9},
  9046\relax
\mciteBstWouldAddEndPuncttrue
\mciteSetBstMidEndSepPunct{\mcitedefaultmidpunct}
{\mcitedefaultendpunct}{\mcitedefaultseppunct}\relax
\EndOfBibitem
\bibitem[Subramaniam \emph{et~al.}(2005)Subramaniam, Abkarian, and
  Stone]{naturec_subramanian_2005}
A.~Subramaniam, M.~Abkarian and H.~Stone, \emph{Nature Materials}, 2005,
  \textbf{4}, 553--556\relax
\mciteBstWouldAddEndPuncttrue
\mciteSetBstMidEndSepPunct{\mcitedefaultmidpunct}
{\mcitedefaultendpunct}{\mcitedefaultseppunct}\relax
\EndOfBibitem
\bibitem[Kotula and Anna(2012)]{sm_kotula_2012}
A.~Kotula and S.~Anna, \emph{Soft Matter}, 2012, \textbf{8}, 10759\relax
\mciteBstWouldAddEndPuncttrue
\mciteSetBstMidEndSepPunct{\mcitedefaultmidpunct}
{\mcitedefaultendpunct}{\mcitedefaultseppunct}\relax
\EndOfBibitem
\bibitem[Park \emph{et~al.}(2009)Park, Nie, Kumachev, Abdelrahman, Binks,
  Stone, and Kumacheva]{a_park_2009}
J.~Park, Z.~Nie, A.~Kumachev, A.~Abdelrahman, B.~Binks, H.~Stone and
  E.~Kumacheva, \emph{Angewandte Chem. Int. Ed.}, 2009, \textbf{48},
  5300--5304\relax
\mciteBstWouldAddEndPuncttrue
\mciteSetBstMidEndSepPunct{\mcitedefaultmidpunct}
{\mcitedefaultendpunct}{\mcitedefaultseppunct}\relax
\EndOfBibitem
\bibitem[Tumarkin \emph{et~al.}(2011)Tumarkin, Park, Nie, and
  Kumacheva]{c_tumarkin_2011}
E.~Tumarkin, J.~Park, Z.~Nie and E.~Kumacheva, \emph{Chem. Commun.}, 2011,
  12712--12714\relax
\mciteBstWouldAddEndPuncttrue
\mciteSetBstMidEndSepPunct{\mcitedefaultmidpunct}
{\mcitedefaultendpunct}{\mcitedefaultseppunct}\relax
\EndOfBibitem
\bibitem[Zoueshtiagh \emph{et~al.}(2014)Zoueshtiagh, Baudoin, and
  Guerrin]{sm_zoueshtiagh_2014}
F.~Zoueshtiagh, M.~Baudoin and D.~Guerrin, \emph{Soft Matter}, 2014,
  \textbf{10}, 9403\relax
\mciteBstWouldAddEndPuncttrue
\mciteSetBstMidEndSepPunct{\mcitedefaultmidpunct}
{\mcitedefaultendpunct}{\mcitedefaultseppunct}\relax
\EndOfBibitem
\bibitem[Bihi \emph{et~al.}(2016)Bihi, Baudoin, Butler, Faille, and
  Zoueshtiagh]{prl_bihi_2016}
I.~Bihi, M.~Baudoin, J.~Butler, C.~Faille and F.~Zoueshtiagh, \emph{Phys. Rev.
  Lett.}, 2016, \textbf{117}, 034501\relax
\mciteBstWouldAddEndPuncttrue
\mciteSetBstMidEndSepPunct{\mcitedefaultmidpunct}
{\mcitedefaultendpunct}{\mcitedefaultseppunct}\relax
\EndOfBibitem
\bibitem[Subramaniam \emph{et~al.}(2006)Subramaniam, Mejean, Abkarian, and
  Stone]{l_subramaniam_2006}
A.~Subramaniam, C.~Mejean, M.~Abkarian and H.~Stone, \emph{Langmuir}, 2006,
  \textbf{22}, 5986--5990\relax
\mciteBstWouldAddEndPuncttrue
\mciteSetBstMidEndSepPunct{\mcitedefaultmidpunct}
{\mcitedefaultendpunct}{\mcitedefaultseppunct}\relax
\EndOfBibitem
\bibitem[Taccoen \emph{et~al.}(2016)Taccoen, Lequeux, Gunes, and
  Baroud]{prx_taccoen_2016}
N.~Taccoen, F.~Lequeux, D.~Gunes and C.~Baroud, \emph{Phys. Rev. X}, 2016,
  \textbf{6}, 011010\relax
\mciteBstWouldAddEndPuncttrue
\mciteSetBstMidEndSepPunct{\mcitedefaultmidpunct}
{\mcitedefaultendpunct}{\mcitedefaultseppunct}\relax
\EndOfBibitem
\bibitem[Poulichet and Garbin()]{pnas_poulichet_2015}
V.~Poulichet and V.~Garbin, \emph{Proc. Nat. Ac. Sci.}\relax
\mciteBstWouldAddEndPunctfalse
\mciteSetBstMidEndSepPunct{\mcitedefaultmidpunct}
{}{\mcitedefaultseppunct}\relax
\EndOfBibitem
\bibitem[Poulichet \emph{et~al.}(2016)Poulichet, Huerre, and
  Garbin]{sm_poulichet_2016}
V.~Poulichet, A.~Huerre and V.~Garbin, \emph{Soft Matter}, 2016, \textbf{13},
  125\relax
\mciteBstWouldAddEndPuncttrue
\mciteSetBstMidEndSepPunct{\mcitedefaultmidpunct}
{\mcitedefaultendpunct}{\mcitedefaultseppunct}\relax
\EndOfBibitem
\bibitem[Strutt (Lord~Rayleigh)(1917)]{pm_strutt_1917}
J.~Strutt (Lord~Rayleigh), \emph{Philos. Mag.}, 1917, \textbf{34}, 94--98\relax
\mciteBstWouldAddEndPuncttrue
\mciteSetBstMidEndSepPunct{\mcitedefaultmidpunct}
{\mcitedefaultendpunct}{\mcitedefaultseppunct}\relax
\EndOfBibitem
\bibitem[Minnaert(1933)]{pm_minnaert_1933}
M.~Minnaert, \emph{Phyl. Mag.}, 1933,  235--248\relax
\mciteBstWouldAddEndPuncttrue
\mciteSetBstMidEndSepPunct{\mcitedefaultmidpunct}
{\mcitedefaultendpunct}{\mcitedefaultseppunct}\relax
\EndOfBibitem
\bibitem[Plesset(1949)]{asme_plesset_1949}
M.~Plesset, \emph{ASME J. Appl. Mech.}, 1949, \textbf{16}, 228--231\relax
\mciteBstWouldAddEndPuncttrue
\mciteSetBstMidEndSepPunct{\mcitedefaultmidpunct}
{\mcitedefaultendpunct}{\mcitedefaultseppunct}\relax
\EndOfBibitem
\bibitem[Keller and Kolodner(1956)]{jap_keller_1956}
J.~Keller and I.~Kolodner, \emph{J. Appl. Phys.}, 1956, \textbf{27},
  1152--1161\relax
\mciteBstWouldAddEndPuncttrue
\mciteSetBstMidEndSepPunct{\mcitedefaultmidpunct}
{\mcitedefaultendpunct}{\mcitedefaultseppunct}\relax
\EndOfBibitem
\bibitem[Brennen(1995)]{oup_brennen_1995}
C.~Brennen, \emph{Cavitation and bubble dynamics}, Oxford University Press, New
  York, 1995\relax
\mciteBstWouldAddEndPuncttrue
\mciteSetBstMidEndSepPunct{\mcitedefaultmidpunct}
{\mcitedefaultendpunct}{\mcitedefaultseppunct}\relax
\EndOfBibitem
\bibitem[Binks(2002)]{cocis_binks_2002}
B.~Binks, \emph{Current Opinion in Colloid and Interface Science}, 2002,
  \textbf{7}, 21--41\relax
\mciteBstWouldAddEndPuncttrue
\mciteSetBstMidEndSepPunct{\mcitedefaultmidpunct}
{\mcitedefaultendpunct}{\mcitedefaultseppunct}\relax
\EndOfBibitem
\end{mcitethebibliography}

\providecommand*{\mcitethebibliography}{\thebibliography}
\csname @ifundefined\endcsname{endmcitethebibliography}
{\let\endmcitethebibliography\endthebibliography}{}

\end{document}